\documentclass[showpacs,preprintnumbers,amsmath,amssymb,prd,aps,
nofootinbib,superscriptaddress,twocolumn,10pt]{revtex4-2}

\usepackage{graphicx}% Include figure files
\usepackage{array}% Include array package
\usepackage{dcolumn}% Align table columns on decimal point
\usepackage{bm}% bold math
\usepackage[columnwise,mathlines]{lineno}
\usepackage{xcolor}% Allow text in colors

\begin{document}

%Preprint number needs a change
%\preprint{\textbf{Draft 3.4, Physical Review D/XXX-GWS}}

\title{Implementation of the rROF denoising method in the cWB 
pipeline for gravitational-wave data analysis}

\author{Pablo J. Barneo}
 \email{pablobarneo@icc.ub.edu}%
 \affiliation{Departament de F\'isica Qu\`antica i
 Astrof\'isica (FQA), Universitat de Barcelona (UB), 
 c. Mart\'i i Franqu\`es 1, E-08028 Barcelona, Spain}%
 \affiliation{Institut de Ci\`encies del Cosmos (ICCUB), 
 Universitat de Barcelona (UB), c. Mart\'i i Franqu\`es 
 1, E-08028 Barcelona, Spain}%
\author{Alejandro Torres-Forn\'e}
 \affiliation{Departamento de Astronom\'ia y Astrof\'isica, 
 Universitat de Val\`encia, Dr.~Moliner 50, 46100 
 Burjassot (Val\`encia), Spain}%
\author{Jos\'e A.~Font}
 \affiliation{Departamento de Astronom\'ia y Astrof\'isica, 
 Universitat de Val\`encia, Dr.~Moliner 50, 46100 
 Burjassot (Val\`encia), Spain}%
 \affiliation{Observatori Astron\`omic, Universitat de
 Val\`encia, Catedr\'atico Jos\'e Beltr\'an 2, 
 46980 Paterna (Val\`encia), Spain}%
\author{Marco Drago}
 \affiliation{Dipartimento di Fisica, Universit\`a di 
 Roma ``La Sapienza'', Piazzale Aldo Moro 2, 
 I-00185 Roma, Italy}%
 \affiliation{INFN, Sezione di Roma, Piazzale Aldo Moro 2,
 I-00185 Roma, Italy}%
\author{Jordi Portell}%
 \affiliation{Departament de F\'isica Qu\`antica i
 Astrof\'isica (FQA), Universitat de Barcelona (UB), 
 c. Mart\'i i Franqu\`es 1, E-08028 Barcelona, Spain}%
 \affiliation{Institut de Ci\`encies del Cosmos (ICCUB), 
 Universitat de Barcelona (UB), c. Mart\'i i Franqu\`es 1, E-08028 Barcelona, Spain}%
 \affiliation{Institut d'Estudis Espacials de Catalunya (IEEC), c. Gran Capit\`a 2-4, E-08034 Barcelona, Spain}%
\author{Antonio Marquina}
 \affiliation{Departamento de Matem\'aticas, Universitat de 
 Val\`encia, Dr.~Moliner 50, 46100 Burjassot (Val\`encia),
 Spain}%

%\collaboration{Virgo Collaboration, LVK Collaboration}

\hspace{1ex}

\date{\today}

\begin{abstract}
The data collected by the current network of gravitational-wave 
detectors are largely dominated by instrumental noise. Total 
variation methods based on $L_1$-norm minimization have 
recently been proposed as a powerful technique for noise 
removal in gravitational-wave data. In particular, the 
regularized Rudin-Osher-Fatemi (rROF) model has proven 
effective to denoise signals embedded in either simulated 
Gaussian noise or actual 
detector noise. Importing the rROF model to existing search 
pipelines seems therefore worth considering. In this paper, we 
discuss the implementation of two variants of the rROF 
algorithm as two separate plug-ins of the coherent Wave Burst (cWB) 
pipeline designed to conduct searches of unmodelled 
gravitational-wave burst sources. The first approach is 
based on a single-step rROF method and the second one employs an 
iterative rROF procedure. Both approaches are calibrated using 
actual gravitational-wave events from the first three observing 
runs of the LIGO-Virgo-KAGRA collaboration, namely GW1501914, 
GW151226, GW170817, and GW190521, encompassing different types 
of compact binary coalescences. Our analysis shows that the 
iterative version of the rROF denoising algorithm implemented 
in the cWB pipeline effectively eliminates noise while preserving 
the waveform signals intact. Therefore, the combined approach 
yields higher signal-to-noise values than those computed by the 
cWB pipeline without the rROF denoising step. The incorporation 
of the iterative rROF algorithm in the cWB pipeline might hence 
impact the detectability capabilities of the pipeline along with
the inference of source properties.
\end{abstract}

\keywords{Gravitational waves detection, gravitational waves 
analysis, LVK, Virgo, denoising, rROF, cWB}%Use showkeys class option
%if keyword display desired
\maketitle

%\tableofcontents

%%%%%%%%%%%%%%%%%%%%%%%%%%%%%%%%%%%%%%%%%%%%%%
\section{\label{sec:introduction}Introduction}
%%%%%%%%%%%%%%%%%%%%%%%%%%%%%%%%%%%%%%%%%%%%%%

The third observational run (O3)~\cite{Abbott2021a} of the 
network of interferometer detectors Advanced 
LIGO~\cite{Aasi2015} and Advanced Virgo~\cite{Acernese2015} has 
led to a significant increase  in the number of transient 
gravitational-wave (GW) detections from compact binary 
coalescences (CBC) with respect to the previous two runs. 
During the first two observing runs (O1 and
O2)~\cite{Abbott2018}, 
the LIGO Scientific Collaboration and the Virgo Collaboration 
reported a total of 11 detections, comprising 10 binary black 
hole (BBH) mergers~\cite{Abbott2016a, Abbott2016b, Abbott2016c}, and 
one binary neutron star (BNS) merger~\cite{Abbott2017}. 
During O3, the number of confirmed detections from CBC events 
climbed to 79, leading to a total of 90 events. Those have been 
recently reported by the LIGO-Virgo-KAGRA (LVK) Collaboration  
in the third release of the Gravitational Wave Transients 
Catalog (GWTC-3~\cite{Abbott2021}).

The notable increase in the amount and variety of waveforms is 
challenging data-analysis procedures. More exceptional events 
are present in the O3 data set~\cite{Abbott2021a, Abbott2021b, Abbott2021c}, 
pushing the capabilities of current data-analysis tools and techniques to 
extract the GW signals embedded in instrumental noise. GW 
searches in the data collected by the interferometers are 
conducted in two different ways: real-time searches using 
low-latency online pipelines, and offline searches on archived 
data. The latter use stand-alone offline versions of the same 
pipelines without any time limitation, allowing for the use of 
higher computational resources.

Real-time searches try to identify event candidates with low 
latency during the observing time, usually less than one minute 
since the detection starts. On the other hand, offline searches 
perform an in-depth analysis of event candidates as well as 
searches for events missed by the speediness of the low-latency 
infrastructure. Both types of analysis require detailed 
background studies, noise characterization and identification, 
and accurate reconstruction of the physical parameters of the 
sources and of their sky localization. Some of this information 
may not be readily available during the low-latency search, 
which makes necessary a posterior offline analysis to recover 
as much information as possible.

GW interferometers work under conditions of low signal-to-noise 
ratio (SNR) and relatively high levels of instrumental noise. 
This makes noise removal (or denoising) one of the most 
challenging problems in GW data analysis. Detector 
characterization techniques have been developed within the LVK 
Collaboration with the purpose of reducing, identifying, and 
characterizing instrumental noise, applying and identifying 
vetoes and gates to the data~\cite{Smith2011, Ajith2014, Aasi2015}. 
Complementary studies on noise reduction using Machine Learning 
methods are currently under intense 
scrutiny~\cite{BAHAADINI2018,TorresForne2020,Cuoco2020,Yu2021}.

In~\cite{Torres2014} methods for denoising GW signals based on 
$L_1$-norm minimization, modelling the denoising problem as a 
variational problem, were first discussed. Originally, these 
methods were developed in the context of image processing where 
they proved to be the best approach to solve the 
Rudin-Osher-Fatemi (ROF) denoising model~\cite{ROF1992}. From 
its original formulation~\cite{ROF1992}, the ROF model has been 
extended to incorporate different denoising alternatives. 
One of these is the {\it regularized} ROF (rROF) denoising 
method whose performance with GW data has been assessed 
in~\cite{Torres2014, Torres2016,Torres2016b,Torres2018}. These 
studies have shown that the rROF method is suitable to denoise 
GW signals embedded either in additive Gaussian 
noise~\cite{Torres2014} or in actual detector 
noise~\cite{Torres2016}, irrespective of the signal morphology
or astrophysical origin of the data. Moreover, it has also been 
found that the rROF method leads to suitable results almost 
irrespective of the data conditioning, the whitening, or the 
removal of spectral artefacts usually present in the analysis 
procedure of GW data analysis pipelines. 

This paper further extends those studies by discussing the 
implementation and calibration of the rROF method in a GW 
data-analysis pipeline, with the goal to make it available in 
upcoming data-taking runs. The selected pipeline is coherent 
Wave Burst (cWB) which is  designed for GW data analysis of 
unmodelled sources~\cite{Klimenko2016,Drago2021}. 
By looking for excess energy on pixels in time-frequency 
representations of the data, cWB is able to identify coherently 
GW transients on a network of GW detectors with minimal 
assumptions on signal morphology. We show here that an 
implementation of the rROF method based on iterative 
regularization~\cite{Osher2005,Xu2013} 
yields satisfactory denoising capabilities when applied to 
actual GW data. 

This paper is organized as follows:  In 
Section~\ref{sec:rROF-method} we briefly describe the rROF 
denoising method and assess its performance through the tuning 
of its intrinsic parameters. Results and procedures for its use 
are also presented in this section. In 
Section~\ref{sec:cWB-pipeline} we discuss our specific 
implementation of the rROF method in the cWB data-analysis 
pipeline. The results of our combined approach are presented in 
Section~\ref{sec:results} using first  signal GW150914 as a 
real-case test and then extending the study to additional GW 
events from O1 to O3. Finally, in Section~\ref{sec:outlook} we 
draw our conclusions and outline possible extensions of this 
work.

%%%%%%%%%%%%%%%%%%%%%%%%%%%%%%%%%%%%%%%%%%%%%%%%
\section{\label{sec:rROF-method}The \lowercase{r}ROF Method}
%%%%%%%%%%%%%%%%%%%%%%%%%%%%%%%%%%%%%%%%%%%%%%%%

%%%%%%%%%%%%
\subsection{\label{subsec:rROF-basics}Basics}
%%%%%%%%%%%%

The starting point of signal denoising is the computation 
of the metric distance between the true (noiseless) signal and the 
noisy signal. In a metric space, this distance is usually 
defined as the square of the $L_2$-norm of the difference of 
both functions, which should be identical to the standard 
deviation of the noise, $\sigma$,
\begin{equation}
   ||{u} - {f}||_{L_2} = \sigma
   \label{eq:sigma-noise}
\end{equation}
where ${f}$ is the observed signal, and ${u}$ is the (unknown) 
signal to be recovered. As usual, we will employ the linear 
degradation model,
\begin{equation}
   {f} = {u} + {n}
\label{eq:def-signal}    
\end{equation}
where ${n}$ represents the additive noise.

To solve the mathematical problem of denoising 
Eq.~(\ref{eq:sigma-noise}), the first approach one can use is 
classical least-squares methods. These methods solve a linear 
system of equations using a linear combination of polynomials 
or wavelets~\cite{Irani1993}, with unknown coefficients. By 
determining those coefficients the denoising problem is solved, 
although the results may be affected by ringing or smearing 
edges effects, known as Gibbs' phenomena~\cite{Marquina2008}. 
In addition, if the linear system is large compared to the size of 
the data sample, finding the solution with least-squares 
methods can be computationally very expensive. 

One of the most common ways to avoid these problems is to 
regularize the least-squares approach, adding an auxiliary 
energy term $R(u)$ to the equation. We will refer to it as the 
regularisation term. This function can be regarded as an {\it a 
priori} probability density. A solution for one-dimensional 
signals, such as a time series, can be found by solving the 
constrained variational problem that results from the addition 
of the regularisation term to Eq.~(\ref{eq:sigma-noise}) (the 
constraint). This problem has a unique solution provided the 
energy function $R(u)$ is convex. Moreover, the variational 
problem can be formulated as an unconstrained variational 
problem using Tikhonov regularisation which adds the constraint 
weighted by a positive Lagrange multiplier $\lambda >0$ to the 
energy
\begin{equation}
    u = 
    \underset{u} {\text{argmin}}
    \left\{R(u) + \frac{\lambda}{2} F(u) \right\}\,.
\label{eq:Ru-definition}    
\end{equation}
Here $F$ is the fidelity term that measures the similarity of 
the solution to the data. This formulation ensures that for a 
positive non-vanishing value of $\lambda$, to be determined, 
there is a unique solution $u$ that matches the constraint. The 
scale parameter $\lambda$ controls the relative significance of 
the fidelity term.

The choice of the energy term $R(u)$ will determine the 
complexity of the problem as well as the properties of the 
solution. For example, if we choose
\begin{equation}
    R(u)= \int||\nabla u ||_{L_2}^2\,,  
\label{eq:Ru-choice}    
\end{equation}
where $\nabla$ stands for the gradient operator, we will obtain 
the so-called Wiener filter. In order to compute the solution 
we solve the associated Euler-Lagrange equation 
\begin{equation}
   \Delta u + \lambda (f-u) = 0\,,
\label{eq:euler-lagrange}    
\end{equation}
under homogeneous Neumann boundary 
conditions~\cite{Evans1991, John1982}. 
Eq.~(\ref{eq:euler-lagrange}) is a non-degenerate second-order, 
linear, elliptic differential equation, which is not difficult 
to solve due to the differentiability and strict convexity of 
the energy term.

Eq.~(\ref{eq:euler-lagrange}) can be solved in an efficient way 
using the Fast Fourier Transform (FFT), which provides a unique 
solution. The Fourier coefficients of the solution decay to 
zero, while those representing the wave $u$ remain with finite 
values. This is no longer the case when the signal contains 
noise because it amplifies high frequencies and yields 
solutions with spurious oscillations near steep gradients or edges. 

The ROF model~\cite{ROF1992} tries to address the problems of 
least-squares methods by replacing the $L_2$-norm in the energy 
term with the $L_1$-norm. By doing this, 
Eq.~(\ref{eq:Ru-definition}) reads
\begin{equation}
    u = \underset{u} {\text{argmin}} \left\{ \int |\nabla u|  + 
    \frac{\lambda}{2}||u-f||^2_{L_2} \right\}\,,
\label{eq:rROF}    
\end{equation}
where the fidelity term is chosen to be equal to the variance 
of the noise $\sigma^2$,
\begin{equation}
    F(u) = ||u-f||^2_{L_2}\,.
\label{eq:F-term}    
\end{equation}
This change  allows recovering edges of the original signal by 
removing noise and avoiding ringing and spurious oscillations. 
Since the energy term $R(u)=|\nabla u|$, called the 
total-variation (TV) norm, is convex, there is a unique optimal 
value of the Lagrange multiplier $\lambda$ for which  
Eq.~(\ref{eq:sigma-noise}) is satisfied.
When the standard deviation of the noise is unknown a heuristic 
estimation of such optimal value is needed. For large enough 
values of $\lambda$ the ROF model will remove very little noise 
while smaller values will have the opposite effect.

However, in the associated Euler-Lagrange equation of the ROF 
model,
\begin{equation}
   \nabla \cdot \frac{\nabla u}{|\nabla u|} + \lambda (f-u) = 0\,,
\label{eq:euler-lagrange-rof}    
\end{equation}
the differential operator becomes singular when $|\nabla u|=0$ 
and has to be defined properly. The algorithm we consider in 
our study is the so-called regularised ROF algorithm 
(rROF)~\cite{Vogel1996}. This algorithm computes an approximate 
solution of the ROF model by smoothing the TV energy. Since the 
Euler-Lagrange derivative of the TV-norm is not well defined 
where $\nabla u= 0$, the TV functional of the rROF method is 
slightly perturbed by introducing in the formulation a small 
positive parameter, $\beta$, 
\begin{equation}
    u = \underset{u} {\text{argmin}} \left\{ \int 
    \sqrt{( |\nabla u|^2 + \beta )} + 
    \frac{\lambda}{2}||u-f||^2_{L_2} \right\}\,.
\label{eq:rROF-model}    
\end{equation}
Here, $u \in \mathbb{R}^p$, where $p$ is the dimension of the 
signal. When $\beta$ is small the problem turns nearly 
degenerate and the algorithm becomes slow in flat regions. In 
contrast, when $\beta$ is large, the rROF method cannot 
preserve sharp discontinuities. Assuming homogeneous Neumann boundary 
conditions, Eq.~(\ref{eq:rROF-model}) becomes a non-degenerate 
second-order nonlinear elliptic differential equation whose 
solution is smooth. To solve Eq.~(\ref{eq:rROF-model}) we use 
conservative, second-order,  central differences for the 
differential operator and point values for the source term. The 
approximate solution is obtained by employing a non-linear 
Gauss-Seidel iterative procedure that uses as an initial guess 
the observed signal $f$. This algorithm has interesting 
properties including robustness and fast convergence.

%%%%%%%%%%%%
\subsection{\label{subsec:param-estimation}Parameter selection}
%%%%%%%%%%%%

The rROF method contains several specifiable parameters. The 
results of the denoising procedure strongly depend on the 
evaluation of these parameters, most importantly on the scale 
parameter $\lambda$ \cite{Torres2017}. As discussed, the optimal 
value of $\lambda$ and of any other parameter in the method, 
cannot be set up a priory. These values must be found 
empirically.
In~\cite{Torres2017} only the scale parameter $\lambda$ was 
evaluated in the calibration of the method. In the present 
investigation, we gauge the values of all algorithm parameters, 
which we shall now describe. The goal is to find a small 
span of parameter values that provide a recovered (denoised) signal for 
all waveforms under different SNR conditions. Parameter $\beta$ 
is needed to avoid divisions by zero in the formulation, which 
implies that the typical values of this parameter will be close 
to zero. Parameter $h$ is inherited from the original ROF model 
proposed for digital image processing and corresponds to the 
step in the finite-difference scheme used to compute the 
gradient. In this context, the value of $h$ should be equal to 
the distance between two adjacent pixels of the image to be 
denoised. However, when adapting the rROF method to GW 
analysis, there is no obvious counterpart explanation about 
the role of $h$. Therefore, we treat $h$ as one more free 
parameter to adjust.

The solution of Eq.~(\ref{eq:rROF-model}) is found through a 
Gauss-Seidel iterative procedure that terminates upon the 
fulfilment of a given condition. In our case, the error of the 
TV minimisation is compared to a control tolerance value 
({\tt tol}), which is an additional parameter to adjust. As we 
discuss below, the correct adjustment of the tolerance plays a 
significant role, as the minimisation process may diverge in 
some situations. 

Finally, to process the data, the entire segment of data 
must be divided into smaller samples of equal size. Each of these 
samples is treated mathematically as the elements of a vector 
with dimension $N$, where $N$ is equivalent to the sample size. 
To optimise the performance of the rROF algorithm we treat $N$ 
as another tunable parameter. We will show that it plays only a 
minor role in denoising. However, it is the most important 
parameter in terms of the computer workload, regarding memory and 
execution speed. The higher the value of $N$ the more computer 
memory is needed and the longer the time the evaluation of the 
parameters takes.

The proper adjustment of these five parameters, $h$, $\beta$, 
$\lambda$, {\tt tol}, and $N$, determines the efficiency 
and the performance of the rROF method when denoising a data segment. 
Our goal will be to find the optimal parameter set, able to 
diminish the amplitude of the noise as much as possible while 
preserving the original signal intact. 
Inadequate selection of parameters can either result in 
insufficient noise removal or in a very aggressive denoising, 
the latter reducing in the process the amplitude of the actual 
GW signal.

To select the parameters we use a hyper-parameter tuning system 
(as described in Section~\ref{subsec:denoising-estimators}). We 
vary the values of the five-parameter set within predetermined 
intervals and perform data denoising for each point in the 
hyper-parameter space. Then, the result is compared to a 
reference data set, usually consisting of a pre-calculated 
waveform template. The hyper-parameter point that provides the 
best denoising result with respect to the reference set will 
allow us to know the optimal parameter values. In our approach, 
a sample of interferometer noise strain and a GW signal are 
needed. Different noise strain samples with different 
characteristics may need a different set of parameter values. 
For this reason, we distinguish between different kinds of 
noisy data by considering, on the one hand, the observational 
run they belong to (O1, O2, and O3) and, on the other hand, 
the interferometer that recorded the noise (H1, L1, or V1).  

%%%%%%%%%%%%%%%%%%%%%%%%%%%%%%%%%%%%%%%%%%%%%%%%%%%%%%
\subsection{\label{subsec:iter-regular}Iterative rROF}
%%%%%%%%%%%%%%%%%%%%%%%%%%%%%%%%%%%%%%%%%%%%%%%%%%%%%%

Through the application of the rROF algorithm, we can compute a 
residual $ v \equiv u - f $. This residual is treated in the ROF 
model as an error and discarded. However, in actual applications 
there will always exist some amount of 
signal in $v$ and some quantity of noise in $u$. The distribution 
depends on the scale parameter $\lambda$. A large value of $\lambda$ 
yields very little noise 
removal, and hence $u$ is close to $f$. On the contrary, a small 
value of 
$\lambda$ yields a noisy, over-smoothed $u$. If the amount of signal 
in $v$ can 
be considered an insignificant fraction of the noise-free 
signal $u$, the residual can be safely discarded treating 
the signal lost as an affordable error. However, if this is 
not the case, a possibility to improve the denoising 
process is to apply the method once again to a new linear 
degradation model that results from using the residual, 
i.e. $f=u+v$. 

This procedure admits a natural {\it iterative} 
generalization, as proposed in~\cite{Osher2006}, that 
solves the deficiencies of the single-step rROF method. Such proposal 
was first applied in the context of GW denoising 
in~\cite{Torres2014}. Here, we 
follow that same approach and build an iterative rROF algorithm 
which uses the decomposition of the data $f$ into a 
candidate to the true noise-free signal $u$ and a residual $v$. 
Therefore, at each iteration $ I_k = u_{k+1} + v_{k+1}$ where 
$k$ is the iteration index and $ I_k = f + v_k$.
The procedure is as follows: 
\begin{itemize}
    \item Initialization: $u_0 = 0$ and $v_0 = 0$ for $k=0$
    \item For $k = 1, 2,...$: compute $u_{k+1}$ as the 
    minimizer of $I_k$ as obtained from the rROF method
    \item Compute the residual $ v_{k+1} = I_{k} - u_{k+1}$, 
    which represents the difference between the input and the 
    output data of the rROF algorithm
    \item Add to the initial noisy data the residual, i.e., $ 
    I_{k+1} = f + v_{k+1} $
\end{itemize}
The iterative regularization adds the ``noise" computed by the 
rROF procedure, $v_1$, back to $f$, the original noisy data. 
Then the sum is processed by the rROF minimization algorithm to 
proceed with the next iteration. The procedure stops when some 
discrepancy principle is satisfied, namely when the square of 
the $L_2$-norm of the residual matches the noise level, 
$||u_k - f||_{L_2} \leq \delta$.

In practice, however, this level may not  be known and it 
becomes necessary to resort to some other termination 
criterion. In~\cite{Osher2006} it was shown that the 
residual decreases monotonically until a stopping index $k$ 
is reached. Should the iterations not be stopped properly, 
the process would converge to the noisy data $f$ and the TV 
of the denoised signal might become unbounded. Thus, our 
iterative rROF algorithm proceeds until the result gets noisier, 
i.e.~until $u_{k+1}$ becomes more noisy than $u_k$. When this 
happens $||v_k||_{L_2}$ has reached its minimum value. The 
iterative procedure is therefore terminated at some index $k$ 
for which the local extrema of the denoised signal do not 
start losing total variation.

The heuristic determination of the index to stop the iterations 
depends on $\lambda$ which is the most important parameter of 
the method. For a large value of $\lambda$ the termination 
criterion may already be satisfied after the first step, which 
would result in a suboptimal reconstruction. This does not 
happen if $\lambda$ is sufficiently small which guarantees that 
the data contained in $u_k$ becomes gradually less noisy until 
the termination index is found. This is the reason why the 
parameter values to use with the iterative regularization 
procedure should be higher than those identified as the 
optimal ones. 

%\note{New paragraphs}
The iterative rROF method, thus,   
profits from the denosing properties of the single-step rROF 
algorithm by slowly denoising the data while 
constantly checking for any signal removal, instead of 
extracting as much noise as possible in only one execution.
Therefore, the parameter values to employ should 
be higher than the optimal ones to slow down the pace of the 
denoising. The single-step rROF algorithm is still in use at each 
step of the iterative method, the main difference residing in the 
manipulation 
of the data through the iterative process, where signal loss is 
avoided by enhancing the portion of data where 
a single-step rROF denoising might fail.
%\note{End new paragraphs}

%%%%%%%%%%%%
\subsection{\label{subsec:denoising-estimators}Denoising estimator}
%%%%%%%%%%%%

To assess the quality of the denoising, an estimator that 
compares the results in every point in the hyper-parameter 
space to reference templates must be used. The estimator we choose is 
known as the first Wasserstein distance~\cite{Dobrushin1970, Wass2020}, 
$\mathcal{W}_1$ (WD in the following). This estimator is a 
distance metric with a finite (bounded) value and it has been 
properly defined to be used with time series. The WD reads
\begin{eqnarray}
    \mathcal{W}_1 = \int_{t_1}^{t_2} |f(t) - u(t)|\,dt\,.
\label{eq:wd-definition}
\end{eqnarray}
There is extensive literature describing its properties as well 
as its relation with other metrics through the corresponding 
transformation rules~\cite{Mariucci2017}. The WD is defined to 
be positive in real space. When it is equal to zero, 
the data sample and the reference are identical. In this way, 
when using this estimator in our hyper-dimensional system, the 
adjustment of the parameter values of the rROF method reduces 
to a minimization problem, where we look for the minimum value 
of $\mathcal{W}_1$. This value will correspond to the optimal 
set of rROF parameters.

For the implementation of the  rROF method in the cWB pipeline, 
we perform the denoising of the GW strain data acquired by 
each interferometer before these data are supplied to the pipeline. 
Using the WD estimator we find that the values of the 
parameters may differ significantly for different 
interferometers depending on their particular (time-dependent) 
noise characteristics or on the template  used. Therefore, in 
order to compare estimator results between different 
interferometers and normalize them, we define the Wasserstein 
scale (WS). When there is no noise present in the data and the 
template is a perfect match of the signal the WS 
will measure 0, which is identical to the value of the WD in 
this situation. On the other hand, when no denoise has been 
performed on the data, the WS will measure 100. In this way, 
the WS is by definition in the interval $[0, 100]$ and can
be considered equivalent to the percentage of noise left 
in the strain data. 

A light-weight software package has been developed for the 
tuning (parameter estimation) of the rROF algorithm. It moves 
over the hyper-parameter space in an automated way to apply the 
rROF algorithm to a data sample. The quality of the results is 
estimated by comparing each outcome with a selected reference 
template using the WS. 
Following~\cite{Torres2014, Torres2016, Torres2016b, Torres2018} 
early tests were performed during the development of the code 
using numerical-relativity waveform templates from both CCSN 
and BBH mergers as reference. Those revealed important information 
about the values of the parameters of the method: (1) their 
ranges are limited in all cases to a small interval; (2) the WS 
shows a characteristic behaviour as a function of each one of 
the parameters; (3) the lower the values of $h$, $\beta$, 
$\lambda$ and {\tt tol}  the better the denoising quality, 
up to some minimum values; (4) parameter $N$ behaves in the opposite 
way showing a plateau at a characteristic value; (5) 
parameter {\tt tol} is related to the order of magnitude of the 
GW strain being denoised. The scan of {\tt tol} may sometimes 
reach a minimum value that can lead to divergences in the 
iterative single-step rROF method.

In our practical application of the iterative procedure we take 
as starting point the results of the rROF parameter estimation 
multiplied by some arbitrary factor. This ensures the use of 
parameters with higher values than the optimal ones.

%%%%%%%%%%%%%%%%%%%%%%%%%%%%%%%%%%%%%%%%%%%%%%
\section{\label{sec:cWB-pipeline}cWB Pipeline}
%%%%%%%%%%%%%%%%%%%%%%%%%%%%%%%%%%%%%%%%%%%%%%

The central goal of this investigation is the implementation of 
the rROF denoising method in the cWB  data-analysis 
pipeline~\cite{Klimenko2016,Drago2021}. The cWB pipeline is 
especially suited for searches of unmodeled GW sources. 
Since no prior information about the morphology of the signal 
is required, cWB can facilitate the detection of GW events for 
which templates cannot be numerically generated or simulated. 
We briefly describe next the basic features of this pipeline.

%%%%%%%%%%%%
\subsection{\label{subsec:cWB-desc}Basics of the cWB pipeline}
%%%%%%%%%%%%

Data analysis from a detector network can be performed using a 
coherent approach, requiring a coincidence in a time window for 
the events identified by the individual detectors, and with 
similar signal morphology. To estimate the statistical 
significance or false alarm rate (FAR) of a GW candidate, the  
responses of individual interferometers in the network are 
compared against the distribution of the expected background. 
By repeating the analysis on many chunks of data, introducing 
non-physical time shifts, allows to invalidate the coherence in 
the data that is exclusively due to random coincidences. 
Therefore, this method allows discriminating between detector 
noise and real signals present in the data. Background 
distributions generated by this time-shifting technique include 
non-Gaussian noise and non-stationary structures in the data.

\newpage
The cWB pipeline~\footnote{cWB home page: 
https://gwburst.gitlab.io/,\\public repositories: 
https://gitlab.com/gwburst/,\\public documentation:\\ 
https://gwburst.gitlab.io/documentation/latest/html/index.html} 
is based on an algorithm that searches for a coherent maximum 
likelihood in the whitened time-series data of the detector 
network employing Wilson-Daubechies-Meyer (WDM) 
transformations. 
This procedure is applied to a multi-resolution time-frequency 
(TF) representation of the data. A more complete representation 
of the data is then obtained using a linear combination of 
wavelet sets at different resolutions.
Triggers are identified by clustering spectrogram pixels over 
the threshold of excess power over the whole interferometer 
network. Then a cluster of pixels is selected, and the 
likelihood statistics are built. 
The cWB pipeline is also able to choose a selection of clusters 
with a given pattern, particularly with a frequency increase as 
a function of time, which is especially suitable to identify 
the inspiral GW signal of compact binary coalescences. The 
statistics of a cWB event are proportional to the coherent SNR 
across the detector network. It also estimates the network 
correlation coefficient,  defined as the ratio between the 
coherent energy and the total energy. This coefficient is 
expected to be close to one for real GW events, and almost zero 
for non-stationary noise fluctuations.

%%%%%%%%%%%%
\subsection{\label{subsec:cWB-implementation}Implementation of 
the rROF method in the cWB pipeline} 
%%%%%%%%%%%%

%% Figure 1
\begin{figure*}[t]
\includegraphics[width=0.47\textwidth]{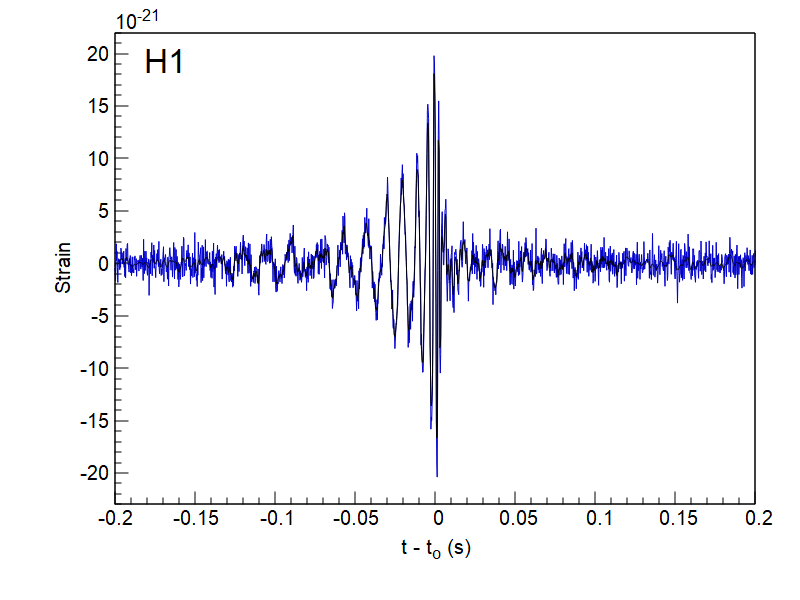}\quad\quad%
\includegraphics[width=0.47\textwidth]{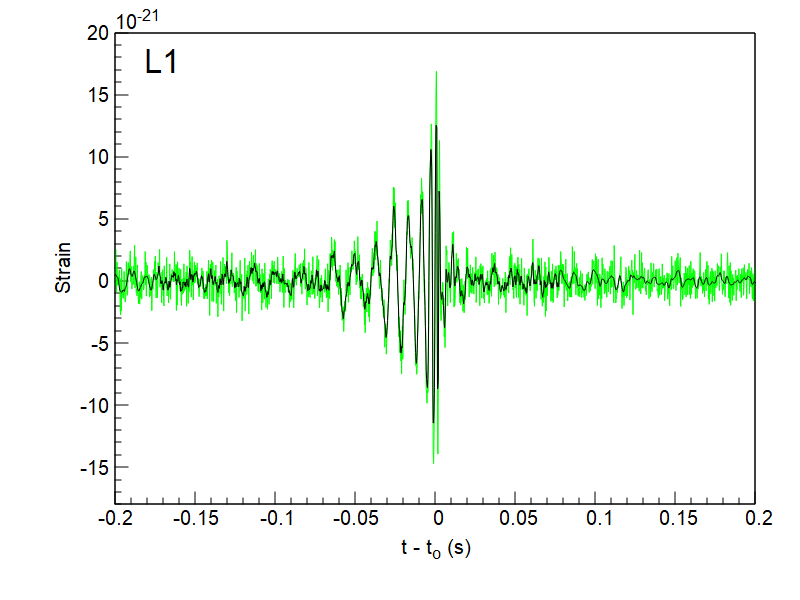}
\caption{\label{fig:denoise-waves}
Comparison between the GW150914 whitened and denoised strains 
for both H1 (left panel) and L1 (right panel). The denoised 
strains  (black lines) are obtained after the application 
of the rROF method to the whitened strain data, using the optimal set 
of parameter values of Table~\ref{tab:gw-par-tuning}. Whitened 
strains are shown in blue (H1) and in green (L1).}
\end{figure*}
%% End figure 1

Data analysis with the cWB pipeline starts first with the 
data-conditioning step. This is done utilizing a regression 
algorithm~\cite{Tiwari2015} that identifies and removes 
persistent lines and noise artefacts. Afterwards, the data is 
whitened and converted to the TF domain using the WDM wavelet 
transformation~\cite{Necula2012}. This analysis is repeated 
several times at several frequency resolutions to obtain 
good TF coverage for a broad range of signal morphologies. Candidate 
events can be identified as a cluster of TF data samples with 
power above the baseline detector noise. In the final step, the 
pipeline reconstructs the signal waveforms, the wave 
polarization and the source sky localization using 
a constrained maximum likelihood analysis over the GW detector 
network~\cite{Drago2021, Klimenko2016}.

The cWB pipeline is written in C++ and is used in combination 
with several ROOT macros. The main functions of the pipeline 
manage the external ROOT macros to use them for specific tasks 
to perform the cWB analysis. This structure allows the 
possibility of adding external routines of any kind, called 
plugins, for any specific purpose that can be combined with the 
default analysis procedure of the pipeline. The implementation 
of the rROF algorithm in the cWB pipeline, both using its 
original design as well as the iterative regularization 
extension, has been developed as plugins. A first plugin was 
built for the single-step rROF method. This routine operates 
over the data stream after the whitening step, which is 
performed by the pipeline itself. The integration at this point 
of the analysis procedure ensures that the application of the 
rROF algorithm is independent of the frequency range of the 
data, as well as of the parameters intrinsic to the algorithm. 
A second plugin has also been developed for the iterative rROF 
algorithm. When used, this second plugin operates in 
replacement of the rROF plugin in the cWB pipeline 
under the same conditions.

%% Figure 2
%\begin{figure*}[t]
%\includegraphics[width=0.47\textwidth]{Figure2-left-panel.png}\quad
%\includegraphics[width=0.47\textwidth]{Figure2-right-panel.png}
%\caption{\label{fig:waveforms}
%GW150914 waveforms obtained by the cWB data-analysis pipeline. 
%Red lines correspond to the cWB-only results while black lines 
%represent the cWB+rROF combined results. H1 data are shown on 
%the left panel and L1 data on the right one.}
%\end{figure*}
%% End figure 2

%% Figure 3
\begin{figure*}[htb]
\centering
\includegraphics[width=0.47\textwidth]{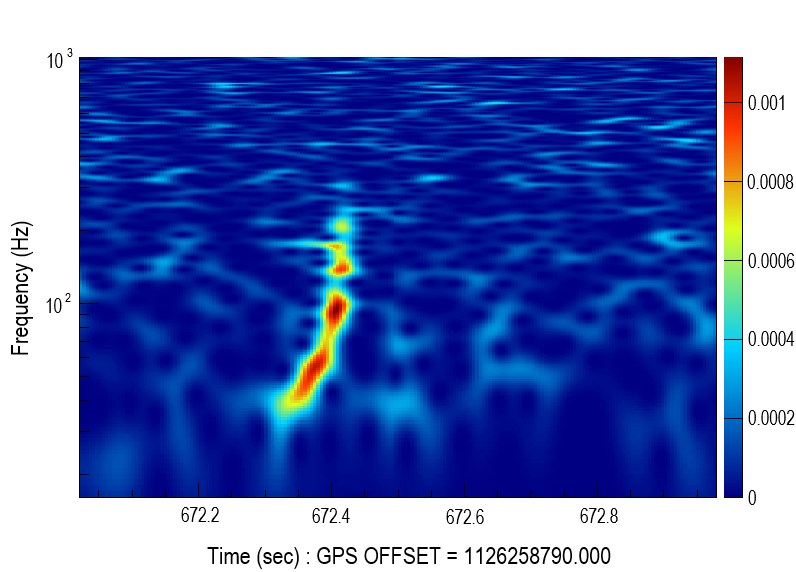} \quad
\includegraphics[width=0.47\textwidth]{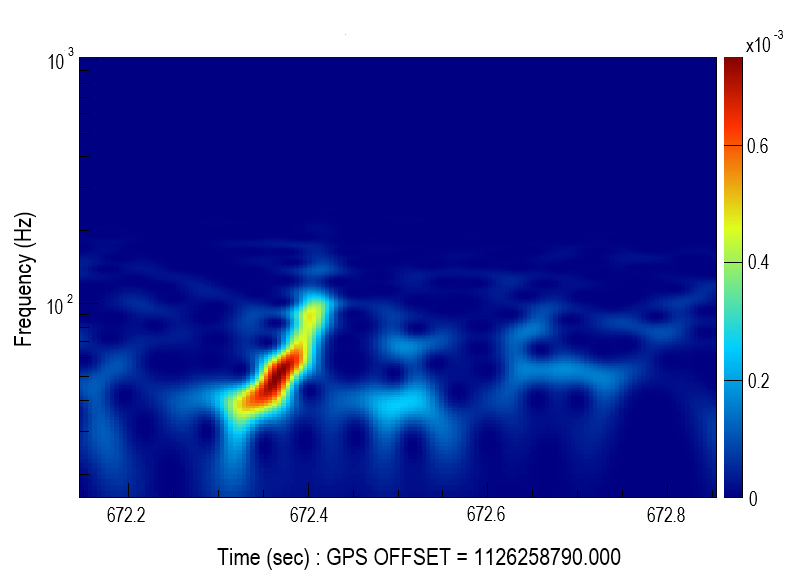} \\ \vspace{1mm}
\includegraphics[width=0.47\textwidth]{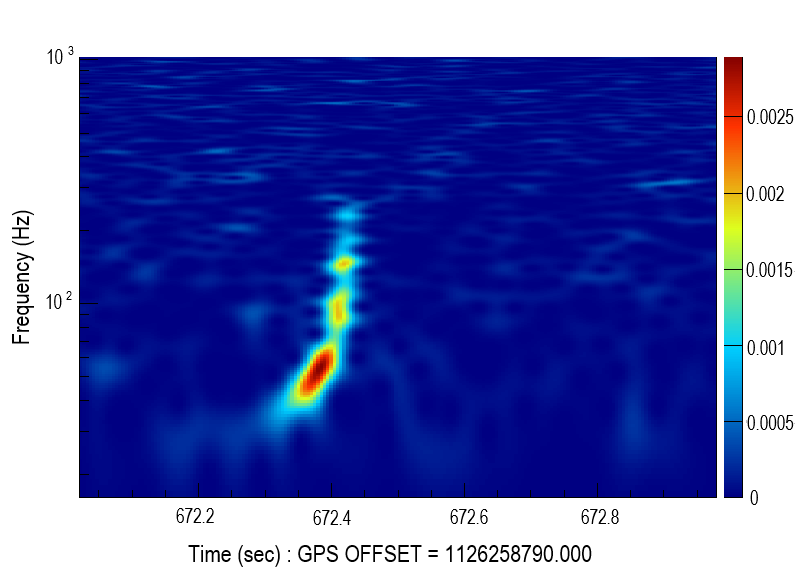} \quad
\includegraphics[width=0.47\textwidth]{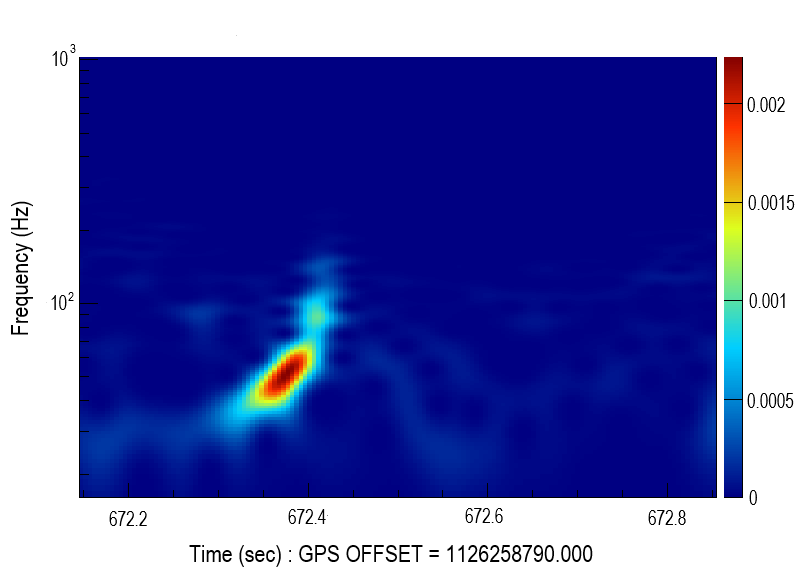}
\caption{\label{fig:espectrograms}
cWB spectrograms for the GW150914 waveform. The left 
spectrograms are cWB-only results while the right ones 
correspond to the combined cWB+rROF results. Data on the top 
panels are for L1 while those on the bottom panels are for H1.}
\end{figure*}
%% End figure 3

%% Figure 4
%\begin{figure*}[htb]
%\centering
%\includegraphics[width=0.47\textwidth]{Figure4-left-panel.png} \quad
%\includegraphics[width=0.47\textwidth]{Figure4-right-panel.png}
%\caption{\label{fig:likelihoods}GW150914 likelihoods 
%computed by 
%the cWB pipeline. The left panel shows the result without the 
%activation of the rROF method while the right panel shows the 
%corresponding result with the single-step rROF method active.}
%\end{figure*}
%% End figure 4

%%%%%%%%%%%%%%%%%%%%%%%%%%%%%%%%%%%%
\section{\label{sec:results}Results}
%%%%%%%%%%%%%%%%%%%%%%%%%%%%%%%%%%%%

To test the implementation and performance of the 
rROF denoising method in the cWB pipeline we employ real GW 
strain data freely accessible through the Gravitational Wave Open Science 
Center~\cite{GWOSC}. The signals we select are two O1 
detections, GW150914~\cite{Abbott2016a} and 
GW151226~\cite{Abbott2016b}, the BNS merger event in O2 
GW170817~\cite{Abbott2017}, and the intermediate-mass 
black hole event in O3 GW190521~\cite{GW190521}. 
Most of the following discussion is focused in GW150914 
which we take as an illustrative example to assess the method. The 
evaluation procedure is as follows: first, we determine the 
optimal parameter values of the original rROF method for the 
GW150914 event; next, we perform the data analysis with the cWB 
pipeline equipped with the rROF denoising method; finally, we 
compare these results with those the cWB pipeline yields when 
the rROF denoising substep is not operational. The same 
approach is then repeated for the iterative rROF algorithm.

%%%%%%%%%%%%
\subsection{\label{subsec:par-tuning}Selection of rROF parameters for GW150914}
%%%%%%%%%%%%

Since GW150914 was observed by the two Advanced LIGO 
interferometers, two sets of rROF parameter values  need to be 
determined, one for each detector. With this purpose, we 
use the BBH waveforms computed by the cWB pipeline as the reference 
template to tune the parameters required by the rROF method. 
Table~\ref{tab:gw-par-tuning} reports the optimal set of 
parameter values we obtain for GW150914.

\begin{table}[t]
\caption{\label{tab:gw-par-tuning}
Optimal parameter values of the GW150914 event obtained 
with the rROF algorithm. Results of the WS are 
also shown in the last column.}
\begin{ruledtabular}
\begin{tabular}{ccccccc}
Detector &$h$ &$\beta$ &$\lambda$ & {\tt tol} & $N$ & WS \\
\hline
L1 & 0.3 & 0.5 & 0.02 & 0.2 & 1024 & 30 \\
H1 & 0.1 & 0.5 & 0.01 & 0.2 & 512 & 31 \\
\end{tabular}
\end{ruledtabular}
\end{table}

The strain data is extracted from the cWB pipeline after 
data-conditioning and whitening.  
Figure~\ref{fig:denoise-waves} 
shows the denoised waveforms for GW150914 with the optimal 
parameter values listed in Table~\ref{tab:gw-par-tuning}. The 
black lines represent the denoised data from the whitened 
strains for both H1 (blue line, left panel) and L1 (green line, 
right panel). As this figure shows, the morphology of the 
reference template waveform is properly preserved after the 
denoising. The evaluation of the rROF method is measured with 
the WS estimator. We find that about 70\% of the original noise 
contained in the signal is subtracted in the case of L1 data 
(69\% for H1 data) while the waveform is preserved quite 
accurately.

%%% \note{New paragraph}
We note that the strain data shown in Figure~\ref{fig:denoise-waves} 
is obtained 
directly from the cWB pipeline right after the whitening process, the 
last step of the data-conditioning stage. This stage includes all 
quality controls, vetoes and removal of potential glitches.
As a result, the waveforms show some modifications with respect to 
the original GW150914 raw-strain data plotted in~\cite{Abbott2016a}. 
Our focus in Figure~\ref{fig:denoise-waves} is to highlight the 
difference in the whitened strain when the denoising rROF step is 
applied or otherwise. 
%%% \note{End new paragraph}

%%%%%%%%%%%%
\subsection{\label{subsec:gw150914-analysis}Combined analysis of GW150914}
%%%%%%%%%%%%

We now reanalyse GW150914 with the active implementation of the 
rROF method in the cWB pipeline, using the optimal 
parameters of Table~\ref{tab:gw-par-tuning}. The cWB data 
analysis reported a successful identification and wave 
reconstruction of the GW150914 event for both H1 and L1. 
The original (cWB only) and the reconstructed (cWB+rROF) spectrograms
for the two interferometers are shown 
in Figure~\ref{fig:espectrograms}. The left panels display the 
original cWB results and the right panels the results obtained 
with the addition of the rROF step (L1 is shown on the top and 
H1 on the bottom). The overall reduction of the noise contained 
in the data is visible in the right plots, providing a clearer 
view of the GW150914 chirp signal. However, the average 
amplitude of the event is reduced as well. This is to be 
expected since the detected signal is a combination of the 
actual gravitational waveform and some amount of noise. Further 
inspection of the spectrograms reveals that the rROF step
also causes the high-frequency component of the signal (the 
ringdown part above 150 Hz approximately) not to display as 
prominently in the denoised data as it does in the original cWB 
spectrogram. The visual comparison of the spectrograms shows, 
indeed, that a portion of the signal at the higher frequencies 
is missing in the combined denoised result.

To further quantify the comparison we analyse the output 
of some of the statistical parameters reported by the cWB pipeline.
A selection is shown in Table~\ref{tab:compared-data} including 
the SNR, the  effective correlated amplitude $\rho$, the 
correlation coefficient {\tt cc}, and the network energy 
disbalance {\tt ED}. The effective correlated amplitude is 
obtained from the SNR according to (assuming a network 
correlation near to one):
\begin{eqnarray}
    \rho = \sqrt{ \frac{\Sigma_i {\rm SNR}_i}{2 N_{\rm IFO}}},    
\end{eqnarray}
where $N_{\rm IFO}$ is the number of interferometers active 
during an event and SNR$_i$ is the signal to noise ratio of the 
individual interferometers. We observe that the values of both 
the SNR and $\rho$ are significantly reduced when the rROF 
denoising step is active. This result is unexpected since a 
reduction of the noise present in the data should produce an 
increase of both quantities. On the other hand, the coherence 
coefficient $cc$  increases by $3.2\%$, from 0.93 to 0.96, when 
the rROF step is active. Hence, this coefficient behaves as one 
would expect in the case of  noise reduction from the data.

\begin{table}[t]
\caption{\label{tab:compared-data}
Parameters reported by the cWB pipeline for the analysis of the 
GW150914 event, with and without the activation of the rROF 
algorithm.}
\begin{ruledtabular}
\begin{tabular}{cccccccc}
 &SNR &$\rho$(L1) &$\rho$(H1) &cc &ED \\
\hline
W/o rROF & 25.2 & 16.7 & 16.0 & 0.93 & -0.01 \\
With rROF & 15.5 & 9.8 & 9.5 & 0.96 & -0.05 \\
\end{tabular}
\end{ruledtabular}
\end{table}

%%% Cut out paragraphs
%To understand the behaviour of SNR and $\rho$ we take a closer 
%look to the waveforms plotted in Fig.~\ref{fig:waveforms}. The 
%comparison between the waveforms with and without the denoising 
%step show good agreement at low and medium frequencies while a 
%significant disagreement is found at higher frequencies. 
%Therefore, the inspiral and merger parts of the signal are 
%correctly recovered when the rROF step is active while the 
%ringdown part is less accurate. On the other hand we plot in 
%Figure \ref{fig:likelihoods} the likelihood as obtained by the 
%cWB pipeline with (right panel) and without (left panel) the 
%use of the rROF denoising step. The reduction of the likelihood 
%for frequencies above $\approx 150$ Hz when the rROF step is 
%active is apparent from this figure. This reduction may be the 
%reason for the unexpected result for SNR and $\rho$.
%%% End cut out paragraph

%%%%%%%%%%%%
\subsection{\label{subsec:gw150914-iterative}Analysis of GW150914 with iterative rROF}
%%%%%%%%%%%%

From the results we have just described it becomes clear that 
the application of the single-step rROF method does not improve 
the results of the standalone cWB pipeline, at least for the 
case of the GW150914 waveform. The single-step denoising 
subtracts a significant fraction of the signal at the higher frequencies.

Here we reanalyze this event by combining the cWB pipeline with 
the iterative rROF algorithm as the denoising step, 
as described in Section~\ref{subsec:iter-regular}. 
This method is designed to compensate for the deficiencies of 
the single-step denoising, which occurs when there is  
significant data loss. The values of the parameters of the method 
used in this case are shown in the first row of 
Table~\ref{tab:gw-par-iterative}. These values are determined 
heuristically by increasing the values regarded as optimal for the 
single-step algorithm, determined in Section~\ref{subsec:par-tuning}. 
We note that we do not need to perform a second parametrization as 
the iterative method is based on the successive application of the 
single-step algorithm with sub-optimal parameter values, combined 
with the use of a data quality check after each step until the 
stopping condition is reached. This procedure still uses the same 
low-pass filtering rROF algorithm implemented in the cWB+rROF 
combined analysis. While the algorithm still behaves effectively  
as a low-pass filter the fact of adding back the residual to the 
initial data several times improves significantly the quality of the 
denoising in the high-frequency part of the signal.

%% Figure 5
\begin{figure*}[htb]
\centering
\includegraphics[width=0.47\textwidth]{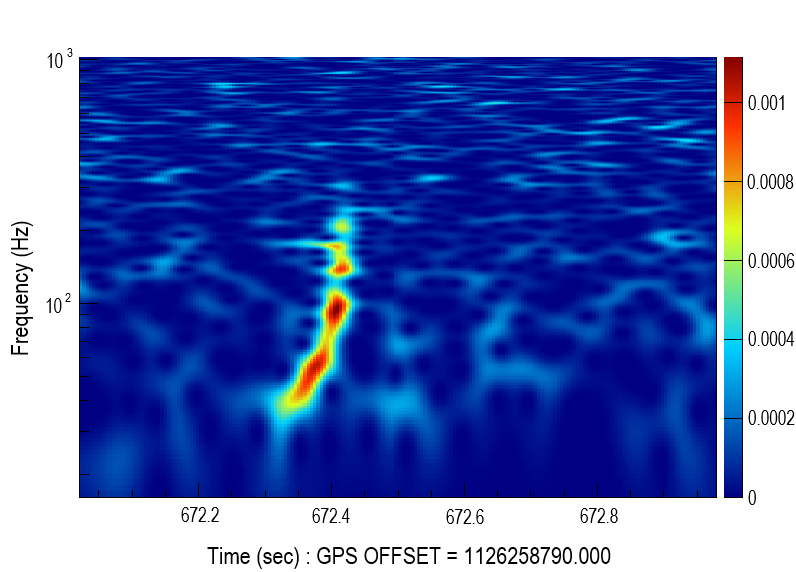} \quad
\includegraphics[width=0.47\textwidth]{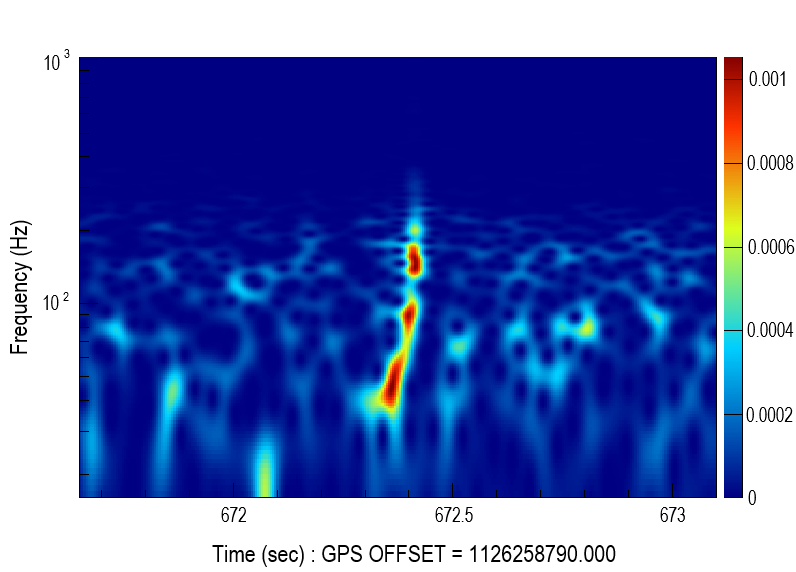} \\ \vspace{1mm}
\includegraphics[width=0.47\textwidth]{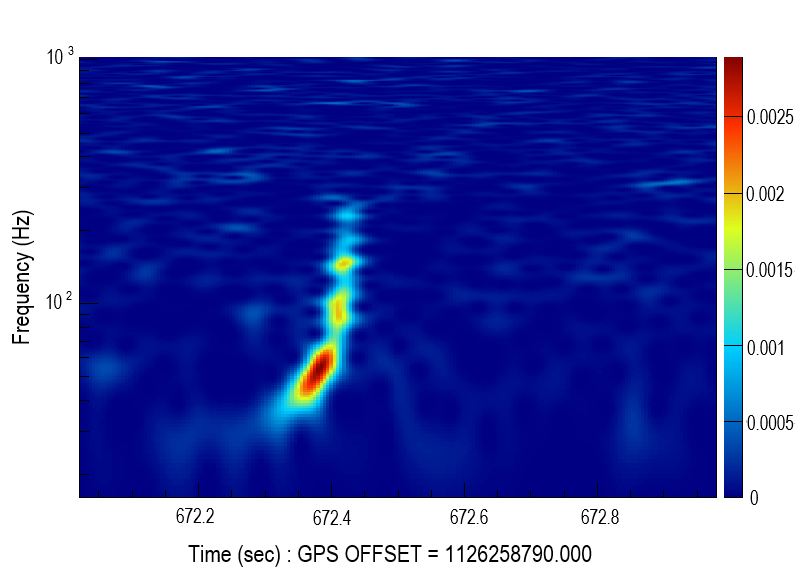} \quad
\includegraphics[width=0.47\textwidth]{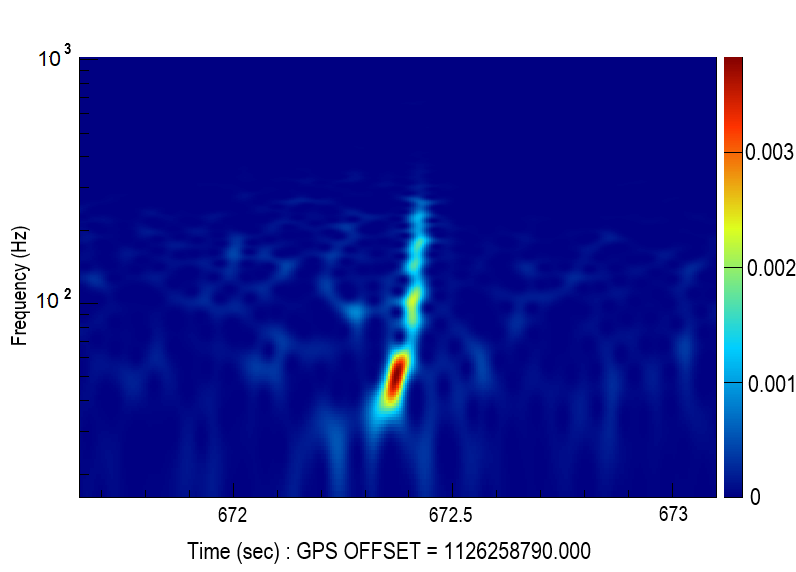}
\caption{\label{fig:espectrograms-iterative}
cWB spectrograms for the GW150914 waveform. The left 
spectrograms are cWB-only results while the right ones 
correspond to the combined cWB and iterative rROF results. Data 
on the top panels are for L1 while those on the bottom panels 
are for H1.}
\end{figure*}
%% End figure 5

\begin{table}[t]
\caption{\label{tab:gw-par-iterative}
Parameter values of the iterative rROF algorithm for the GW 
events considered in this work.}
\begin{ruledtabular}
\begin{tabular}{cccclc}
GW event &$h$ &$\beta$ &$\lambda$ & {\tt tol} & $N$ \\
\hline
GW150914 & 1 & 1 & 0.1 & 0.2       & 1024 \\
GW151226 & 1 & 1 & 0.2 & 0.2       & 1024 \\
GW170817 & 1 & 1 & 1.0 & 0.001     & 1024 \\
GW190521 & 1 & 1 & 0.1 & 0.2       & 1024 \\
\end{tabular}
\end{ruledtabular}
\end{table}

The cWB pipeline reports once again a successful identification 
and waveform reconstruction of the GW150914 event. 
Figure~\ref{fig:espectrograms-iterative} displays the new 
spectrograms obtained from the cWB pipeline for L1 (top panels) 
and H1 (bottom panels). The left column shows the original cWB 
spectrograms without any denoising step active (as in the left 
panels of Fig.~\ref{fig:espectrograms}) and the right column 
the corresponding spectrograms obtained with the combined cWB 
and iterative rROF algorithm.

As for the case of the single-step rROF method, the iterative 
rROF algorithm also yields a visible overall reduction of noise 
which provides a somewhat clearer track of the chirp, specially 
at frequencies higher than $\approx 150$ Hz. The most notable 
difference with respect to the single-step rROF method is that 
the iterative rROF algorithm succeeds in keeping the 
high-frequency part of the signal intact, showing no data loss 
above $\approx 150$ Hz (compare the right panels of 
Fig.~\ref{fig:espectrograms} and 
Fig~\ref{fig:espectrograms-iterative}). Therefore, when 
combining the cWB pipeline with the iterative rROF algorithm 
the complete ringdown part of the GW150914 signal remains 
intact and clearly visible. We also notice that the 
spectrograms of the denoised signals are extremely clean at 
high frequencies (displaying a uniformly dark blue) as was also 
the case when using the single-step rROF method. For the 
latter, that is an indication that the rROF algorithm behaves 
as a low-pass filter. With iterative regularization this is 
still the case since the rROF algorithm is used at every 
iteration. However, by adding the residual back to the signal 
the rROF algorithm behaves as a low-pass filter just for noise, 
thus keeping intact the signal contained in the data. 

When inspecting the numerical values of the statistical 
parameters computed by the cWB pipeline when used in 
combination with the iterative rROF algorithm, we pay special 
attention to the reported SNR as our main indicator of a 
successful denoising. As shown in the first row of 
Table~\ref{tab:events-snr}, the SNR of the GW150914 event 
increases from 25.2 to 27.1, an enhancement of 7.5\% with 
respect to the original cWB measurement. We thus conclude that 
the results obtained with the iterative denoising for our 
selected test case, GW150914, are worth considering.

%% Figure 6
\begin{figure*}[htb]
\centering
\includegraphics[width=0.47\textwidth]{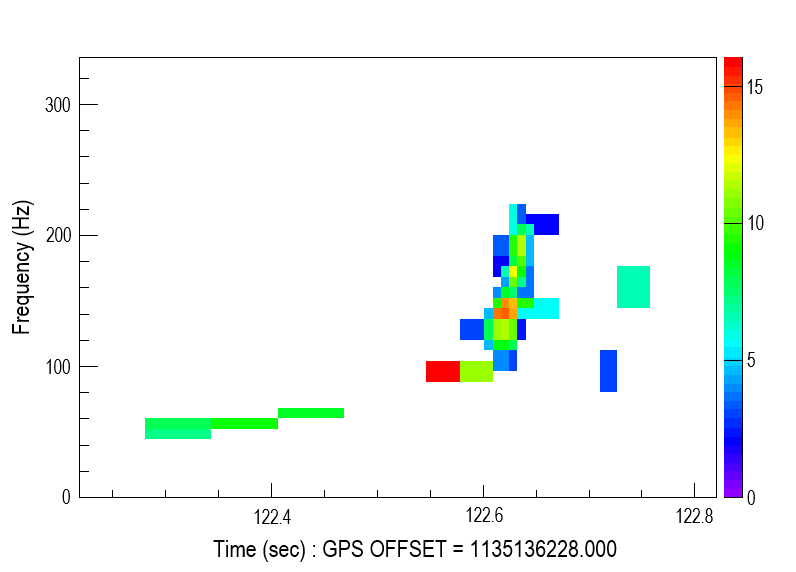} \quad
\includegraphics[width=0.47\textwidth]{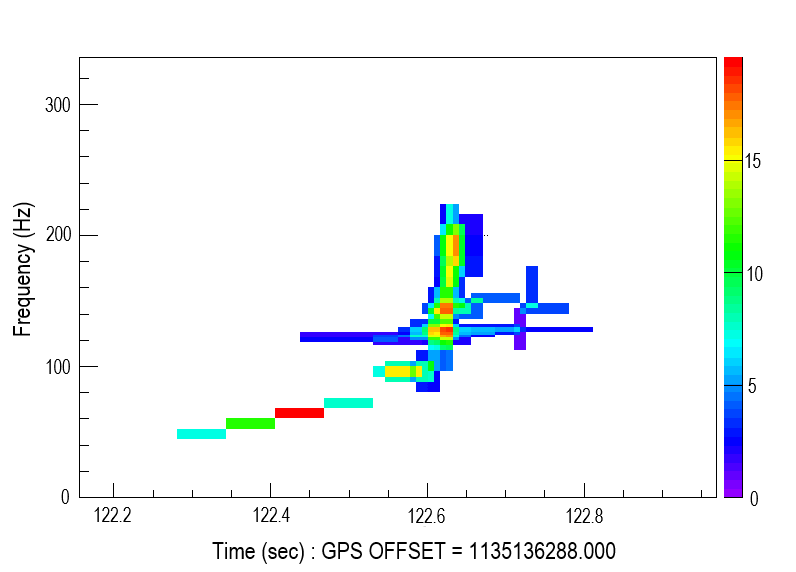} \\ \vspace{1mm}
\includegraphics[width=0.47\textwidth]{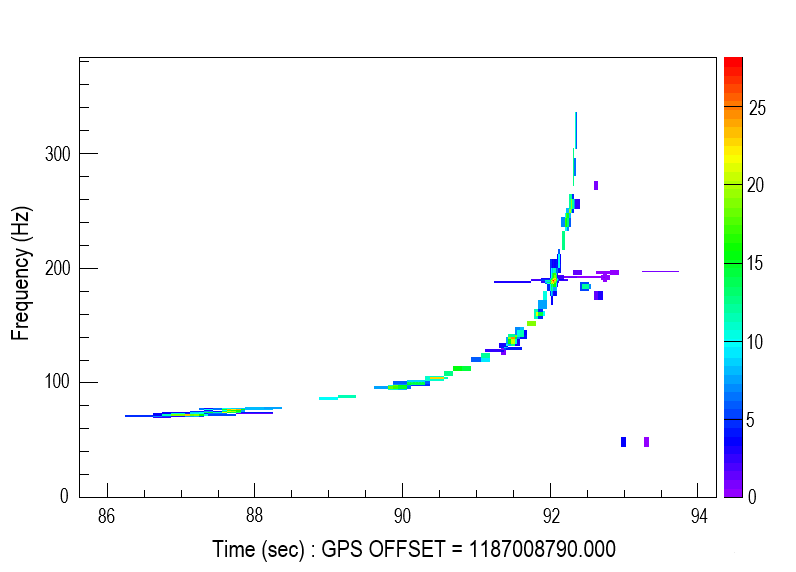} \quad
\includegraphics[width=0.47\textwidth]{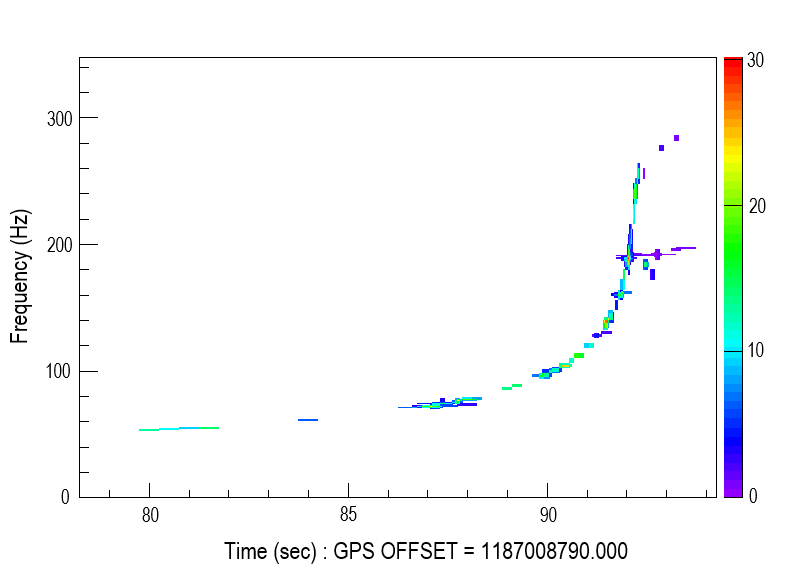} \\ \vspace{1mm}
\includegraphics[width=0.47\textwidth]{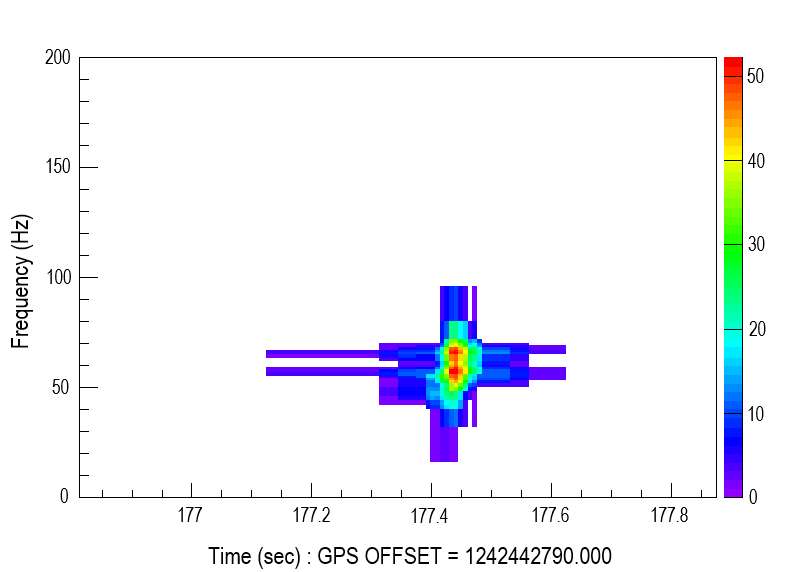} \quad
\includegraphics[width=0.47\textwidth]{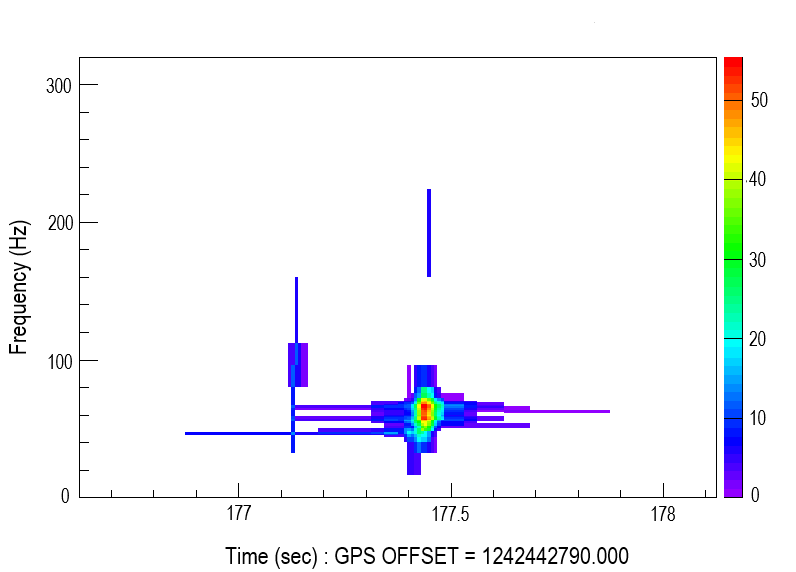}
\caption{\label{fig:likelyhood-gw-events}
Likelihoods computed by the cWB pipeline for GW  events 
GW151226 (top), GW170817 (centre), and GW190521 (bottom). In 
the results displayed in the left column the rROF method is not 
active. The right column shows the corresponding likelihoods 
when cWB is combined with the iterative rROF algorithm.}
\end{figure*}
%% End figure 6

%%%%%%%%%%%%
\subsection{\label{subsec:gwevents-iterative}Additional GW events}
%%%%%%%%%%%%

To complete the assessment of the rROF method as a denoising 
plugin of the cWB pipeline we extend our investigation to 
additional GW events. We aim to prove that the denoising method 
can provide positive results with any signal type regardless of 
the nature of the noise contained in the data. To do this we 
select events GW151226, a BBH merger signal from observing run 
O1~\cite{Abbott2016b}, GW170817, the BNS merger event detected 
in O2~\cite{Abbott2017}, and GW190521, an intermediate-mass 
black hole signal observed in O3~\cite{GW190521}. The 
corresponding SNR values computed by cWB are reported in 
Table~\ref{tab:events-snr}.

We begin by using the cWB pipeline in combination with the 
single-step rROF algorithm. For none of the three events the 
pipeline is able to report a detection. Our conclusion is that 
in all three cases the subtraction of signal during the 
denoising step is more severe than in the case of GW150914, 
despite the fact that we used the optimal parameter values as 
determined for each event separately. 
By removing too much signal from the data the cWB pipeline is 
unable to achieve an identification. Our hypothesis is that it 
might be related to the low-frequency filtering nature of the 
rROF algorithm, which does not perform appropriately for the 
low SNR event GW151226 nor for the high-frequency signal 
GW170817. To obtain a conclusive statement would require a 
deeper analysis of the data subtracted by the rROF denoising.

However, the combined application of cWB and the iterative rROF 
method to the additional GW events yields entirely satisfactory 
results. Using the specific values of the iterative rROF method 
parameters indicated in Table~\ref{tab:gw-par-iterative} we 
find that all three signals are identified by the cWB  
pipeline, the analysis software is able to reconstruct all 
events and in all cases it reports an enhancement in the 
waveform SNR. Table~\ref{tab:events-snr} summarizes the SNR 
values obtained for the four GW events analysed in this work. 
The specific SNR increments are 7.5\% (GW150914), 17.6\% 
(GW151226), 1.1\% (GW170817), and 14.2\% (GW190521).

Figure~\ref{fig:likelyhood-gw-events} displays the likelihood 
computed by cWB for each event: GW151226 in the top row, 
GW170817 in the central row, and GW190521 at the bottom. The 
left column shows the likelihood for each event without the use 
of a rROF denoising step, while the right column displays the 
corresponding likelihood with the iterative rROF algorithm 
active. This figure demonstrates that for all GW events 
considered, the waveforms are identified and properly 
reconstructed by the cWB pipeline. The iterative rROF algorithm 
does not introduce any kind of data loss in any part of the 
spectrograms, in particular in the high-frequency region.  

Finally, we come back to the issue of the parameter values used by 
the iterative regularization, reported in 
Table~\ref{tab:gw-par-iterative}. The methodology indicates that 
values higher than the optimal ones should be used. For all signals 
considered the parameter selection has been made aiming to find an 
acceptable denoising result, taking advantage of the flexibility
that the iterative regularization offers. Moreover, a completely 
operational denoising method should be able to successfully operate 
on any kind of data, without any prior knowledge about the signal. In 
this regard, the possibility of using free parameters independent of 
the kind of noise or signal is in our interest. This is indeed an 
attractive additional possibility iterative regularization offers. 

\begin{table}[ht]
\caption{\label{tab:events-snr}
Values of the SNR computed by cWB for the GW events considered 
in this work. SNR$_a$ corresponds to the purely cWB value (no 
rROF step) while SNR$_b$ is the SNR obtained using cWB in 
combination with the iterative rROF method.}
\begin{ruledtabular}
\begin{tabular}{ccccc}
Event & Type & Run & SNR$_a$ & SNR$_b$ \\
\hline
GW150914 & BBH & O1 & 25.2 & 27.1 \\
GW151226 & BBH & O1 & 11.9 & 14.0 \\
GW170817 & BNS & O2 & 26.8 & 27.1  \\
GW190521 & IMBH & O3 & 14.7 & 16.8  \\
\end{tabular}
\end{ruledtabular}
\end{table}

%%%%%%%%%%%%%%%%%%%%%%%%%%%%%%%%%%%%%%%%
\section{\label{sec:outlook}Conclusions}
%%%%%%%%%%%%%%%%%%%%%%%%%%%%%%%%%%%%%%%%

The rate of detections of GW signals by the Advanced 
LIGO-Virgo-KAGRA interferometer network is dramatically 
increasing with every detector update~\cite{Abbott2021}. The 
data collected are largely dominated by instrumental noise 
which renders data denoising and signal reconstruction truly 
important  efforts. In such context, the denoising of GW 
signals based on $L_1$-norm minimization approaches shows 
strong potential, as evidenced by the results reported 
in~\cite{Torres2014,Torres2016,Torres2016b,Torres2018}. Those 
studies have shown that the regularized Rudin-Osher-Fatemi 
method~\cite{ROF1992} is suitable to denoise GW signals 
embedded either in additive Gaussian noise~\cite{Torres2014} or 
in actual detector noise~\cite{Torres2016}, irrespective of the 
signal morphology, data conditioning, or whitening. 

In this paper, we have extended those studies by discussing the 
implementation and calibration of the rROF method in an 
existing GW data-analysis pipeline, with the mid-term goal of 
having it operational in upcoming LVK data-taking runs. We 
have selected the cWB pipeline, designed for coherent searches 
of unmodelled burst sources~\cite{Klimenko2016,Drago2021} and 
we have implemented the rROF method as a plug-in within the 
flowchart of the pipeline. Additionally, building on a proposal 
laid out in~\cite{Torres2014} we have also implemented an 
iterative regularization approach (as a second plug-in) using 
the single-step rROF algorithm as the base denosing method for 
each iteration. The combined cWB+rROF approach has initially 
been tested using actual noisy data from the GW150914 event. 
The comparison between the results of the cWB pipeline with and 
without the rROF denoising substep has revealed some 
limitations of the single-step rROF method. Our implementation 
of the algorithm has led, in particular, to a significant 
elimination of the high-frequency component of the signal 
(along with the anticipated  noise removal). The remedy to this 
drawback has been found in the iterative rROF algorithm, an 
approach proposed as an improvement over the original model, 
especially formulated to compensate for the signal removal 
sometimes present in a single-step rROF method. The assessment 
of the iterative rROF algorithm with the GW150914 waveform has 
led to satisfactory results. We have found that a notable 
amount of noise can be removed while at the same time the 
entire signal morphology is unaffected 
{\it at all frequencies}, yielding an increment in the cWB 
analysis indicators, most importantly the SNR.

Our analysis has been completed with three additional GW 
events, spanning representative CBC morphologies and detector 
noise, namely a second BBH merger from O1, GW151226, the BNS 
merger event from O2 GW170817, and the intermediate-mass black 
hole event from O3 GW190521. For all of these events we have 
also observed that the iterative version of the rROF algorithm 
implemented in the cWB pipeline leads to an effectual reduction 
of the noise without affecting the signals, thus yielding 
enhanced SNR values. For GW170817 the SNR increment is a 
modest 1.1\% while for GW151226 and GW190521 is 17.6\% 
and 14.2\%, respectively.

As a near-term goal we plan to perform offline analysis of the 
complete O1-O3 data with the combined cWB + iterative rROF 
pipeline. By doing so we will reevaluate the detectability 
capabilities of the pipeline for existing triggers and the 
inference of the source properties, as well as investigate 
whether the improved quality of the data may lead to unveil 
potential new triggers on the available data. In addition, we 
also plan to implement and deploy the iterative rROF method in 
the low-latency version of the cWB pipeline for O4 and O5. The 
experience to be gained with the O1-O3 searches should pave the 
way to the eventual application of the denosing technique 
discussed in this paper to the upcoming observational campaigns 
of the LIGO-Virgo-KAGRA detector network.

\hspace{1ex}

\begin{acknowledgments}
We thank Tom\'as Andrade for his helpful comments and suggestions 
during the writing of this paper, and Giovanni Prodi for a 
careful reading of the manuscript.

This research has made use of data obtained from the 
Gravitational Wave Open Science Center 
(https://www.gw-openscience.org), a service of LIGO Laboratory, 
the LIGO Scientific Collaboration and the Virgo Collaboration. 

This material is based upon work supported by NSF's LIGO 
Laboratory which is a major facility fully funded by the 
National Science Foundation. Virgo is funded, through EGO, by 
the French Centre National de Recherche Scientifique (CNRS),
the Italian Istituto Nazionale della Fisica Nucleare (INFN) and 
the Dutch Nikhef, with contributions by institutions from 
Belgium, Germany, Greece, Hungary, Ireland, Japan, Monaco, 
Poland, Portugal, Spain. 

This work was partially funded by the Spanish 
MICIN/AEI/10.13039/501100011033 and by ``ERDF A way of
making Europe'' by the European Union through grant
RTI2018-095076-B-C21, and by the Institute of Cosmos Sciences
University of Barcelona (ICCUB, Unidad de Excelencia
'Mar\'{\i}a de Maeztu') through grant CEX2019-000918-M. 
PB acknowledges support from the the 
Gestió d'Ajuts Universitaris i de Recerca (AGAUR) by the 
FI-SDUR 00122 (2020). JAF and ATF acknowledge support from the 
Spanish Agencia Estatal de  Investigaci\'on 
(PGC2018-095984-B-I00) and  by  the Generalitat Valenciana 
(PROMETEO/2019/071). MD acknowledges the support from the 
Amaldi Research Center funded by the MIUR program 
'Dipartimento di Eccellenza' (CUP:B81I18001170001) and 
the Sapienza School for Advanced Studies (SSAS). AM 
acknowledges support from the project PID2020-118236GB-I00.
\end{acknowledgments}

%%%%%%%%%%%%%%%%%%%%%%%%%%%%%%%%%%%%%%%%%%%%
%% \bibliography{bibliography}
%%%%%%%%%%%%%%%%%%%%%%%%%%%%%%%%%%%%%%%%%%%%

\providecommand{\noopsort}[1]{}\providecommand{\singleletter}[1]{#1}%
%
%%
%% End bibliography
%%%%%%%%%%%%%%%%%%%%%%%%%%%%%%

\end{document}